\documentclass[10pt,twocolumn,superscriptaddress,longbibliography,floatfix,aps,prx]{revtex4-2}

\usepackage{amsmath, bm, amsfonts, amssymb} 
\usepackage{graphicx}                      
\usepackage{xfrac}                          
\usepackage{grffile}                        
\usepackage{xcolor}
\usepackage[permil]{overpic}
\usepackage{psfrag}
\usepackage{pict2e}
\usepackage{float}
\usepackage{relsize}
\usepackage[caption=false]{subfig}

\newcommand{\aparam}{a}
\newcommand{\xg}{x_{\rm g}}
\newcommand{\sdot}{\dot \sigma}
\newcommand{\gdot}{\dot \gamma}
\newcommand{\gdotbar}{\bar{\dot \gamma}}
\newcommand{\gdotc}{\bar{\dot \gamma}_{\rm c}}

\newcommand{\sigmas}{\sigma_{\rm ss}}
\newcommand{\sigmay}{\sigma_{\rm y}}
\newcommand{\sigmar}{\sigma_{\rm res}}
\newcommand{\ts}{t_{\rm ss}}
\newcommand{\smax}{\sigma_{\rm max}}
\newcommand{\smin}{\sigma_{\rm min}}

\newcommand{\taux}{\tau_x}

\usepackage[colorlinks, citecolor=blue, urlcolor=red, linkcolor=blue]{hyperref}
\begin{document}
\newcommand{\vkw}[1]{\textcolor{black}{#1}}
\newcommand{\smf}[1]{\textcolor{black}{#1}}
\title{Shear banding as a cause of non-monotonic stress relaxation after flow cessation}

\author{Vanessa K. Ward}
\email{vanessa.k.ward@durham.ac.uk}
\author{Suzanne M. Fielding}

\affiliation{Department of Physics, Durham University, South Road, Durham DH1 3LE, United Kingdom}

\date{\today}
\begin{abstract}

Recent flow cessation experiments on soft materials have shown a counter-intuitive non-monotonic relaxation of the shear stress: following the switch-off of a steady imposed shear flow, the stress initially decays before later increasing again. By simulating the soft glassy rheology model in a form extended to allow steady state shear banding, we show that the presence of shear bands prior to flow cessation can give  rise to this phenomenon. We give a mechanistic understanding of the basic physics involved, in terms of (i) the decay of the shear bands after flow cessation, and (ii) the evolution of frustrated local stresses, governed by different time scales for plastic relaxation in each band. In particular, an elastic recoil in the unsheared band gives rise to negative local frustrated stresses, the slow release of which can cause an increase in macroscopic stress. Given that shear banding and frustrated local stresses arise widely across disordered soft solids, we argue that non-monotonic stress relaxation after flow cessation may occur in many different materials. 
\end{abstract}

\maketitle

Soft materials such as  colloids, polymers, emulsions and granular matter display rich rheological behaviour, arising from the interaction of their constituent substructures (polymer chains, emulsion droplets, etc) with an imposed flow~\cite{larson1999structure}. For any such material, a key rheological fingerprint is the ``flow curve'' relating shear stress and shear rate in steady shear, $\sigmas(\gdot)$.  For a broad class of amorphous ``soft glassy materials'', including dense colloids, foams, emulsions, and low density gels, this displays a non-zero yield stress $\sigmay=\lim_{\gdot\to 0}\sigmas$~\cite{Berthier2011,Nicolas2018,bonn2017yield}. When subject to a load $\sigma<\sigmay$, such materials show an elastic solid-like response, or slow creep, with their constituent substructures unable to rearrange significantly. In contrast, larger loads $\sigma>\sigmay$ facilitate rearrangements, leading to yielding and the onset of liquid-like flow.  Significant effort has been devoted to understanding this solid-to-liquid yielding transition, as reviewed in Ref.~\cite{divoux2023ductile}.

In recent years, attention has increasingly turned to understanding the reverse {\em liquid-to-solid} transition that arises on flow cessation. Starting from an initial state of steady shear  with stress $\sigmas(\gdot)$, the shear rate is switched to zero at some time, defined as $t=0$, and the subsequent stress decay observed, $\sigma(t>0)$. Typically, this decay shows a broad spectrum of relaxation timescales, with fast initial decay modes pertaining to the rearrangement of sub-structures on small length-scales, and slower modes signifying larger scale restructuring.  
In soft glasses, the decay is often non-integrable~\cite{Ballauff2013,Moghimi2017,Mohan2013,Vasisht2022,osuji2008shear,negi2010physical,fritschi2014mode},  either with $\sigma\sim t^{-\beta}$ and $0<\beta<1$ or asymptoting to a non-zero constant $\sigmar=\lim_{t\to\infty}\sigma(t)$. The residual stress $\sigmar$ then  provides a macroscopic signature of frustrated stresses that remain trapped locally within the material, as captured in molecular simulations~\cite{Vasisht2022}, elastoplastic models~\cite{Vasisht2022} and mode coupling theory~\cite{fritschi2014mode}. It plays a significant role in controlling mechanical  memory~\cite{Pashine2019,keim2014mechanical,mukherji2019strength,keim2019memory} and performance~\cite{hirobe2021simulation}, as in the spectacular exploding ``Rupert's drop'' phenomenon~\cite{kooij2021explosive}. For a review, see~\cite{withers2007residual}. 

For decades, it has been implicitly  assumed that  the stress must decay monotonically as a function of time $t$ after flow cessation, whether to zero or a non-zero residual value $\sigmar$.  However, recent experiments have challenged this assumption: a remarkable non-monotonic stress decay following flow cessation was reported in boehmite gels~\cite{Sudreau2022} and in three different self-associating organo/supramolecular gels~\cite{Hendricks2019}. Non-monotonic stress evolution has also been seen following  more complicated two-stage compression protocols in granular matter~\cite{Murphy2020}, crumpled sheets~\cite{Lahini2017} and particle simulations~\cite{mandal2021memory}.

To date, such observations remain poorly understood theoretically. Indeed, it is difficult to intuit a mechanism via which stress can increase over any time interval, once shear driving no longer acts on the material. In  associating gels~\cite{Hendricks2019}, it was suggested to arise via the reformation of broken bonds after flow cessation. The released energy was argued to then increase the system's elastic energy via the formation of domains partly aligned with the prior shear. \vkw{Within a phenomenological structural kinetic model~\cite{joshi2022thixotropy}, the stress increase was predicted to arise due to the competition of viscoelastic aging with rejuvenation.}
In the context of constitutive modeling of thixotropy, the buildup of a material's modulus over time has been argued to violate the second law of thermodynamics~\cite{larson2015constitutive}, although elasticity was argued to contribute significantly to thixotropy in Ref.~\cite{choi2021role}.

In this Letter, we put forward an alternative explanation for the phenomenon of non-monotonic stress relaxation after flow cessation. \vkw{In particular, we} argue that the widespread physical ingredients of (i) coexisting shear bands of differing viscosity prior to flow cessation and (ii) frustrated local stresses can interact to give rise to it. We substantiate this scenario by simulating the widely used soft glassy rheology model, in a form extended to allow shear banding. \vkw{It is known that shear banding is present in a number of the systems that exhibit the stress increases~\cite{Vlassopolous.personal}.} We furthermore offer a detailed mechanistic understanding of the basic underlying physics, which can be understood via the evolving distributions of frustrated local stresses in the two former bands, governed by their differing effective temperatures.

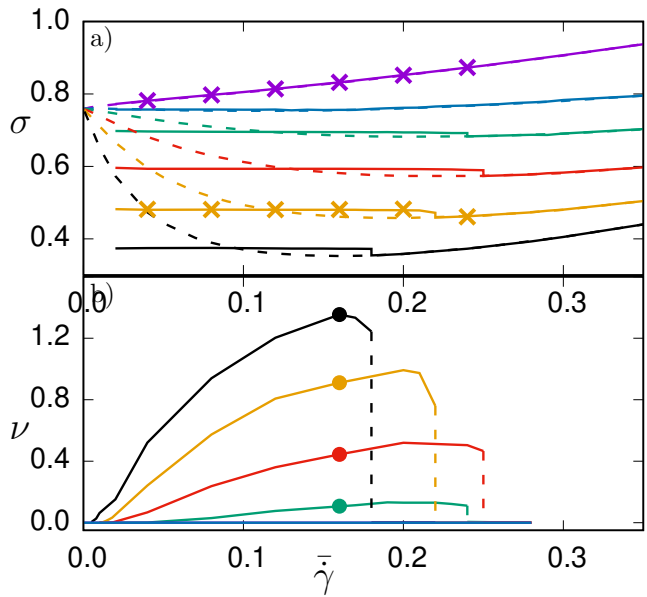
\begin{figure}[!t]
\subfloat{%
\begin{overpic}[width=0.46\textwidth]{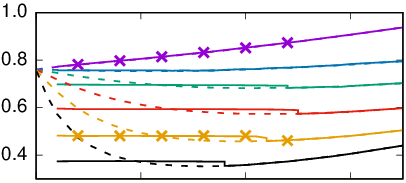}
\put(100,380){\normalsize{a)}}
\put(-30,240){\rotatebox{0}{$\mathlarger{\mathlarger{\mathlarger{\mathlarger{\sigma}}}}$}}%
\end{overpic}
}\vspace{-0.5cm}

\subfloat{%
\begin{overpic}[width=0.46\textwidth]{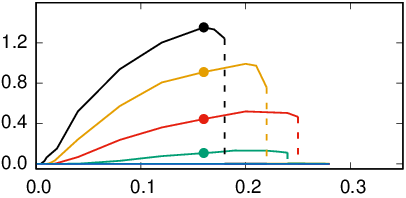}
\put(100,440){\normalsize{b)}}
\put(-30,220){\rotatebox{0}{$\mathlarger{\mathlarger{\mathlarger{\mathlarger{\nu}}}}$}}%
\put(460,-20){$\mathlarger{\mathlarger{\mathlarger{\mathlarger{\bar{\dot \gamma}}}}}$}%
\end{overpic}
}\hfill
 \caption{{\bf a)} Steady state flow curves of shear stress as a function of shear rate (solid lines) and constitutive curves for shear artificially constrained to be homogeneous (dashed lines). From top to bottom, the coupling parameter $\aparam=0.0, 1.0, 2.0, 4.0, 8.0, 16.0$. Crosses indicate shear rates for which the stress decay after flow cessation is explored in Fig.~\ref{fig:relaxation}.  {\bf b)} Non-monotonicity parameter $\nu$, as defined in the main text. The delay time is $\tau_x=100$ and $\aparam$ increases from bottom to top. Dots indicate the systems explored in Fig.~\ref{fig:nonmono}.
}
\label{fig:flowCurves}
\end{figure}

{\it The SGR model ---} considers an ensemble of elastoplastic elements exploring an energy landscape of traps~\cite{Sollich1997,Sollich1998}. Each element  corresponds to a mesoscopic region of material that is large enough to allow the definition of a local continuum shear strain $l$ and shear stress $kl$, with modulus $k$ the same for all elements. The macroscopic elastoplastic stress   is the average of the elemental ones, $\sigma=k\langle l \rangle$.  Under an imposed shear of rate $\dot\gamma$, each element strains at rate $\dot l = \dot\gamma$, leading to an elastic buildup of stress. Elemental stresses are also intermittently released via local plastic events, occurring stochastically at rate $r = \tau_0^{-1} {\rm min} \{1,\exp\left[-\left(E - \frac{1}{2}kl^2\right)/x\right]\}$, with a microscopic attempt time $\tau_0$, trap energy depth $E$, and noise temperature $x$. After yielding, an element resets its local strain $l$ to zero and hops into a new trap of depth drawn from an exponential distribution $\rho\left( E\right) = \exp\left(-E/\xg \right)/\xg$. This gives a glass transition at  $x=\xg$~\cite{Bouchaud1992} and rheological ageing for $x < \xg$~\cite{Fielding2000}.

The SGR model was extended in Ref.~\cite{Fielding2009} to capture a multi-valued constitutive curve $\sigma(\gdot)$ for states of stationary homogeneous shear (dashed lines in Fig.~\ref{fig:flowCurves}). Shear bands can then form when spatial heterogeneity is allowed in the flow-gradient direction $y$, giving a flat plateau in the steady state flow curve (solid lines in Fig.~\ref{fig:flowCurves}). The extended model allows the noise temperature $x=x(y,t)$ to evolve according to
\begin{equation}
\label{eqn:xdynamics}
\taux \frac{\partial x}{\partial t}=-x+x_0+\aparam\langle r l^2\rangle +\lambda^2\frac{\partial^2x}{\partial y^2}.
\end{equation}
Here $x_0$ is a bare noise temperature, $\aparam$ the magnitude of a coupling term representing the pumping of noise by jump events, $\taux$ a delay time for the relaxation of $x$, and $\lambda$ a lengthscale that sets the width of the interface between shear bands. To simulate this model in shear, we place  $M$ SGR elements on each of $S$ streamlines stacked across the flow gradient direction from $y=0$ to $L_y$. 
For an imposed shear rate $\bar{\dot\gamma}\left(t\right)$ averaged across streamlines, the local shear rate $\gdot(y,t)$ in creeping flow follows by setting the total stress $\Sigma = \sigma\left(y,t\right) + \eta\dot\gamma\left(y,t\right) = \bar\sigma\left(t\right) + \eta \bar{\dot\gamma}\left(t\right)$, with $\eta$ a small solvent viscosity, giving 1D Eshelby stress propagation~\cite{picard2004elastic}.  Force balance constrains $\Sigma$ to be uniform across $y$.  

{\it Protocol ---} We simulate a shear protocol as follows.  First, we apply a strong pre-shear of rate $\gdotbar_{\rm pre}$, for a time sufficient to attain a fully fluidised steady state. Second, we drop the shear rate to some value  $\gdotbar$ and hold it constant for a duration $\ts$, sufficient for the stress to attain a steady state prescribed by the flow curve $\sigmas(\gdotbar)$. (To facilitate a more rapid numerical evolution to a banded state, we evolve the system initially with a small $\taux$ before introducing the $\taux$ of interest only once the stress has reached a steady state at $t=-500$.)  Although the macroscopic stress attains a steady state,  the low shear band still slowly ages. The shear is then switched off by setting $\gdotbar=0.0$ at time $t=0$ and the stress relaxation is tracked as function of time post switch-off.

{\it Parameters ---} The parameters of the model are the modulus $k$, microscopic time $\tau_0$, glass transition temperature $x_g$, bare noise temperature $x_0$,  coupling parameter $\aparam$, delay time $\taux$, stress diffusivity $\lambda$ and solvent viscosity $\eta$. The flow device has gap size $L_y$. The parameters of the protocol are the rate  of pre-shear $\gdotbar_{\rm pre}$, the duration $\ts$ of shearing at the shear rate of interest prior to switch-off, and that shear rate $\gdotbar$. Numerical parameters are the number of streamlines $S$ and  elements per streamline $M$, and the timestep $dt$.  We choose units of stress $k=1$, time $\tau_0=1$ and length $L_y=1$. We also set $x_g=1.0$, thereby setting the yield strain $O(1)$. Unless otherwise stated we then set $x_0=0.3$,  $\lambda=0.01$, $\eta=0.05$,  $\gdotbar_{\rm pre}=10.0$, $\ts=1000.0$, $S=100$, $M=1000$ and $dt=0.001$ (increased to $dt=0.01$ for $t>10000$.) Important parameters to explore are the coupling parameter $\aparam$, the imposed shear rate $\gdotbar$, and the delay time $\tau_x$. In particular, $\aparam$ controls the effective temperature and therefore, together with $\gdotbar$, the extent of shear banding prior to flow cessation. The delay time $\tau_x$ controls the relaxation of the effective temperature, and therefore also of the bands, after cessation. 

{\it Results ---} We start in Fig.~\ref{fig:flowCurves}a) by showing the flow curves of shear stress  as a function of shear rate in steady shear prior to switch-off, $\sigma_{\rm ss}(\gdotbar)$.  Solid lines correspond to systems in which the shear field is allowed to become heterogeneous.  Dashed lines show the underlying constitutive curves, with the shear artificially constrained to be uniform. For  $\aparam>\aparam^*\approx 0.84$, the  constitutive curves are non-monotonic, with a regime of negative slope $d\sigma/d\gdot < 0$ that causes shear bands to form over a window of shear rates $0<\gdotbar<\gdotc(\aparam)$. Over this window, the flow curves are flat and differ from the homogeneous constitutive curves. \vkw{The strain rates in the two shear bands that form in this regime correspond to the shear rates at each end of the plateau in the steady state flow curve. The proportion of the gap occupied by each band is such that the shear rate averaged over the two bands, across the gap, equals the applied strain rate. The interface between the two bands has a slightly diffuse width governed by the prefactor $\lambda$ in the final term of Eqn.~\ref{eqn:xdynamics}.} For $\gdotbar>\gdotc$, the shear is again homogeneous, and the constitutive and flow curves coincide. In principle, the flat stress plateau should extend rightwards until it intersects the constitutive curve. In practice, for higher shear rates in this regime the system gets stuck in a state of metastable homogeneous shear, giving a slight step in the flow curve. 

\begin{figure}[!t]
\subfloat{%
\begin{overpic}[width=0.2675\textwidth]{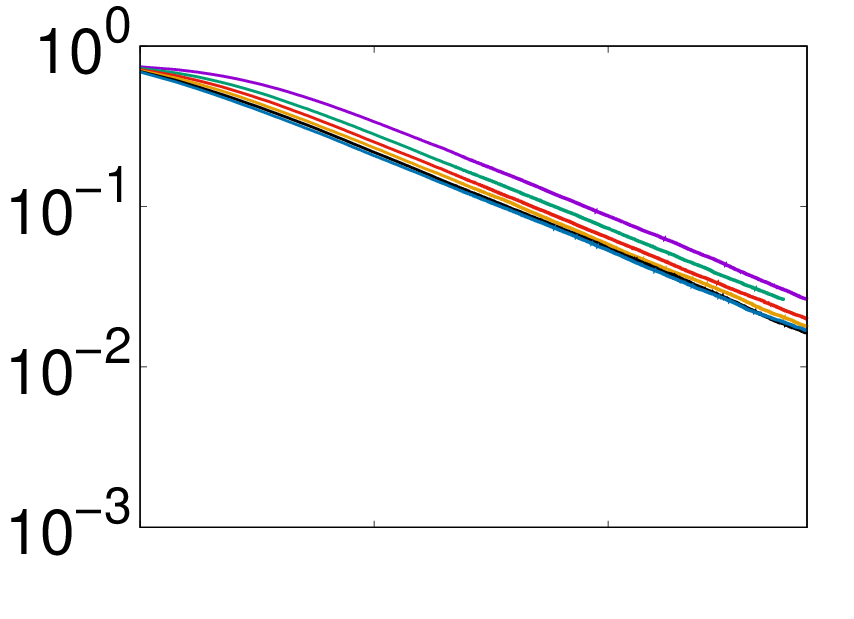}%
\put(300,700){\large{\bf{Homogeneous}}}%
\put(-30,350){$\mathlarger{\mathlarger{\mathlarger{\mathlarger{\sigma}}}}$}
\put(650,600){\normalsize{a) $\aparam=0$}}%
\end{overpic}
}
\subfloat{%
\hspace{-1.26cm}
\begin{overpic}[width=0.2675\textwidth]{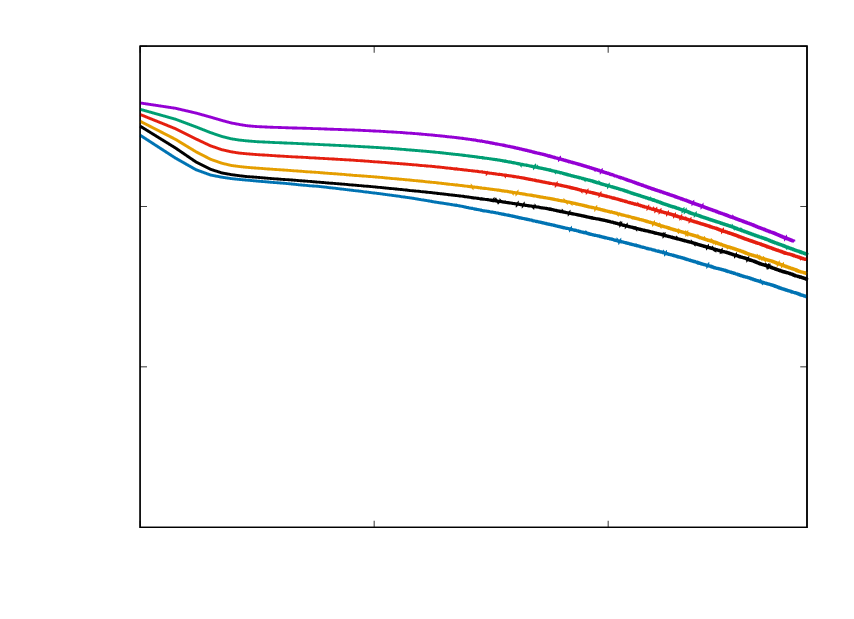}%
\put(300,700){\large{\bf{Shear-banded}}}%
\put(390,600){\normalsize{b) $\aparam=8$, $\tau_x=1$}}%
\end{overpic}
}

\vspace{-1.15cm}
\subfloat{%
\begin{overpic}[width=0.2675\textwidth]{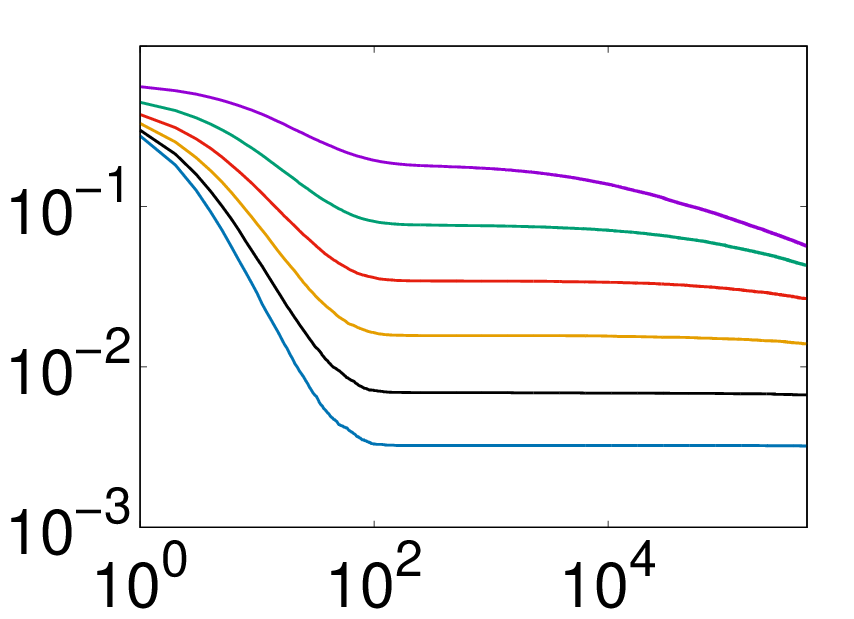}%
\put(330,600){\normalsize{c) $\aparam=8$, $\tau_x=100$}}%
\put(-30,350){$\mathlarger{\mathlarger{\mathlarger{\mathlarger{\sigma}}}}$}
\put(550,-20){$\mathlarger{\mathlarger{\mathlarger{\mathlarger{t}}}}$}
\end{overpic}
}\hfill
\subfloat{%
\hspace{-1.55cm}
\begin{overpic}[width=0.2675\textwidth]{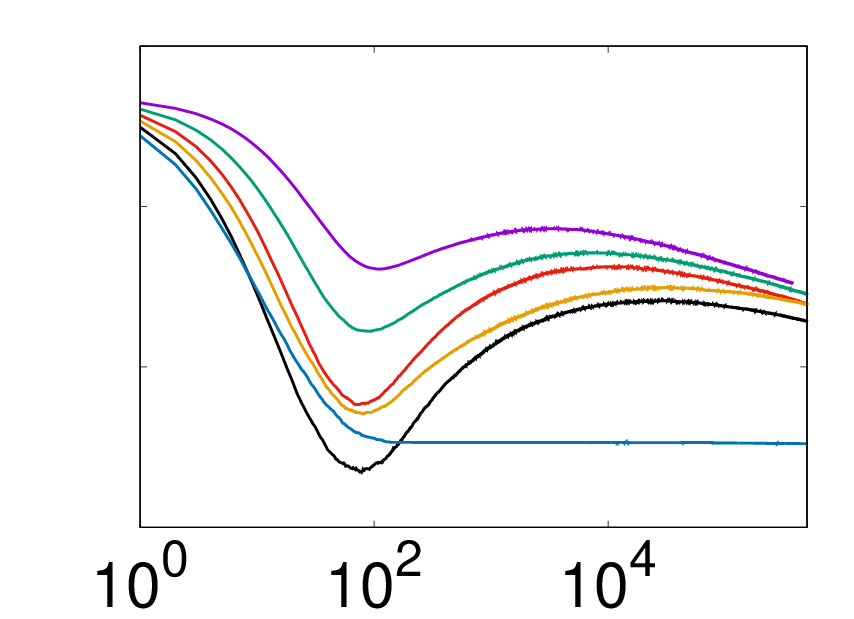}%
\put(550,-20){$\mathlarger{\mathlarger{\mathlarger{\mathlarger{t}}}}$}
\put(320,600){\normalsize{d) $\aparam=8$, $\tau_x=100$}}%
\end{overpic}
}
\caption{Relaxation of the shear stress $\sigma$ as a function of time $t$ after flow cessation. The shear rates  prior to switch-off are $\gdot=0.04,0.08...0.24$ (indicated by crosses in Fig.~\ref{fig:flowCurves}a) in curves from top to bottom in each panel. {\bf a)} Coupling parameter $\aparam=0.0$: homogeneous shear prior to flow cessation. {\bf b)} $\aparam=8.0$ and $\tau_x=1.0$: shear-banded but $x$ decays quickly. {\bf c)} $\aparam=8.0$ and $\tau_x=100.0$: homogeneous shear (artificially constrained.) {\bf d)} $\aparam=8.0$ and $\tau_x=100.0$: shear-banded (except at the largest $\gdotbar=0.24$) and $x$ decays slowly.}
\label{fig:relaxation}
\end{figure}
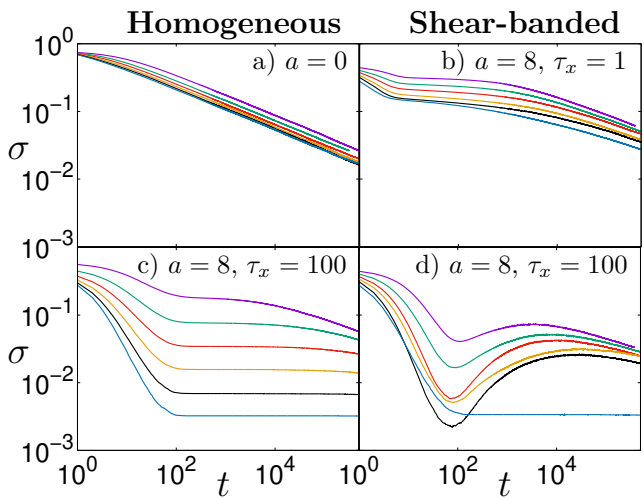

Fig.~\ref{fig:relaxation} shows the relaxation of the shear stress as a function of time $t$ following shear cessation, starting from a state of steady shear prescribed by the flow curves just discussed, such that $\sigma(t=0^+)=\sigmas$. Two values of the coupling parameter $\aparam$ are explored. For $\aparam=0.0$  the underlying constitutive curve in Fig.~\ref{fig:flowCurves} is monotonic, the shear prior to flow cessation is homogeneous, and the stress decay in Fig.~\ref{fig:relaxation}a) is monotonic  for all values of shear rate simulated (indicated by purple crosses in Fig.~\ref{fig:flowCurves}a).

For $\aparam=8.0$, in contrast, the underlying constitutive curve  is non-monotonic and the shear prior to flow cessation is banded for most shear rates indicated by yellow crosses in Fig.~\ref{fig:flowCurves}, in any simulation that allows spatial heterogeneity in the flow-gradient direction $y$. We then find the stress decay after flow cessation to be non-monotonic, provided the delay time $\taux$ is sufficiently large. See Fig.~\ref{fig:relaxation}d). (An exception is seen for the highest shear rate, where the flow prior to switch-off is homogeneous, resulting in monotonic stress decay after switch-off.) In simulations performed for the same parameter values,  but with the shear field artificially constrained to be homogeneous,  the stress decay after cessation is always monotonic (Fig.~\ref{fig:relaxation}c). For the smaller delay time $\taux=1.0$ in Fig.~\ref{fig:relaxation}b), the stress decay after cessation is also monotonic, even with banded flow prior to cessation.  

\begin{figure}[!t]
\subfloat{%
\begin{overpic}[width=0.262\textwidth]{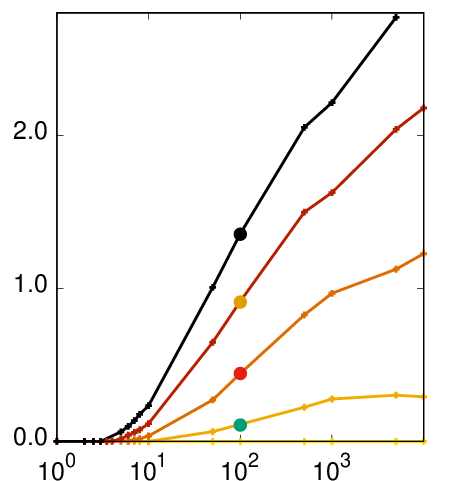}
\put(150,900){\normalsize{a)}}
\put(-40,500){\rotatebox{0}{$\mathlarger{\mathlarger{\mathlarger{\mathlarger{\nu}}}}$}}%
\put(370,-40){$\mathlarger{\mathlarger{\mathlarger{\mathlarger{\tau_x}}}}$}%
\end{overpic}
}\hspace{-1.08cm}
\subfloat{%
\begin{overpic}[width=0.262\textwidth]{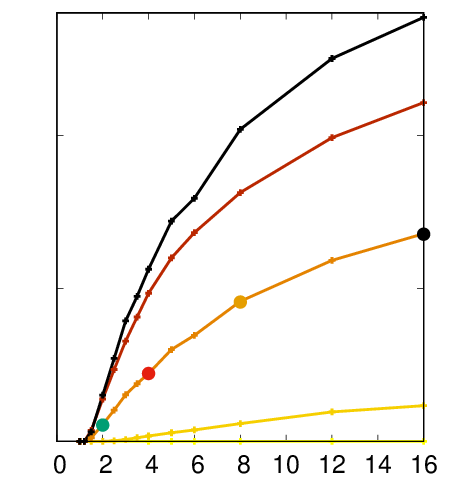}
\put(150,900){\normalsize{b)}}

\put(470,-60){$\mathlarger{\mathlarger{\mathlarger{\mathlarger{\aparam}}}}$}%
\end{overpic}
}
\caption{Non-monotonicity parameter $\nu=\log_{10}(\smax/\smin)$ for $\gdotbar=0.16$. {\bf a)} Plotted vs. delay time $\taux$ for $\aparam=1.0, 2.0,4.0, 8.0, 16.0$ from bottom to top. ($\nu=0$ for $\aparam=1.0$.) {\bf b)}  Plotted vs. coupling parameter $\aparam$ for $\taux=1.0, 10.0, 100.0, 1000.0, 5000.0$ from bottom to top. ($\nu=0$ for $\taux=1.0$.). The dots show parameter values from Fig.~\ref{fig:flowCurves}b.}
\label{fig:nonmono}
\end{figure}
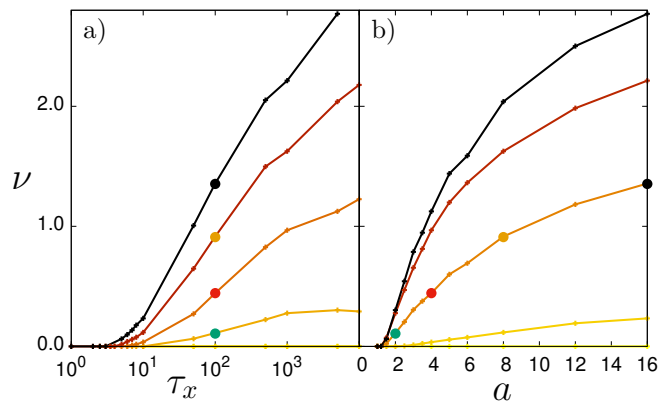

Taken together, the results in Fig.~\ref{fig:relaxation} suggest that two conditions are needed for non-monotonic stress relaxation after flow cessation. The first is the presence of shear bands in the steady state prior to flow cessation. The second is a sufficiently large delay time $\taux$ in the relaxation of the effective temperature after flow cessation -- which, as we shall see below, controls the evolution of the local stress distributions post switch-off. 

To explore further the dependence of this phenomenon on the parameters of the model, we  quantify the degree to which the stress decay after flow cessation is non-monotonic in any simulation via the parameter $\nu=\log_{10}(\smax/\smin)$. Here $\smin$ is the local minimum in the stress relaxation, which occurs around a time $O(10^2)$ for the data in Fig.~\ref{fig:relaxation}d), and $\smax$ the local maximum, around a time $O(10^4$).  A value $\nu=n$ thus corresponds to a stress rise of $n$ orders of magnitude. The values obtained for $\nu$ are averaged over 10 simulation runs.

This non-monotonicity parameter $\nu$ is plotted in Fig.~\ref{fig:flowCurves}b) versus the shear rate prior to flow switch-off, for several values of the coupling parameter $\aparam$ and a delay time $\taux=100.0$. For  $\aparam<\aparam^*$, the shear prior to flow cessation is homogeneous and the stress decay after flow cessation is monotonic  ($\nu=0.0$) at all shear rates. For $\aparam>\aparam^*$ the shear prior to flow cessation is banded for a range of shear rates $0<\gdotbar<\gdotc$ and the stress decay after flow cessation can be non-monotonic, $\nu> 0.0$. In this regime, $\nu$ increases with increasing shear rate $\gdotbar$, and hence with the width of the high shear band. The value of $\nu$ then reaches a peak before declining slightly then falling precipitously to  zero once homogeneous shear is recovered for $\gdotbar\geq\gdotc$.

For one particular imposed shear rate, $\gdot=0.16$, shown by dots in Fig.~\ref{fig:flowCurves}a, we also explore the effect of 
varying the delay time $\taux$ (Fig.~\ref{fig:nonmono}a) and the coupling parameter $\aparam$ (Fig.~\ref{fig:nonmono}b). As can be seen, to obtain a non-monotonic stress decay after flow cessation, $\nu>0$, a sufficiently large value of the coupling parameter is required, $\aparam \geq \aparam^{**} > \aparam^*$. This ensures that the flow state prior to switch-off is shear banded, and with a sufficiently large effective temperature difference between the bands. For $\aparam>\aparam^{**}\approx 1.5$, $\nu$ then increases with increasing $\aparam$.  Also required is a sufficiently long delay time for the relaxation of the effective temperature post switch-off, $\taux \geq \taux^*\approx 5.0$. For $\taux>\taux^*$, $\nu$ then increases with $\taux$. \vkw{The time at which the stress reaches a minimum is largely independent of the imposed shear rate and the coupling parameter $a$, but increases linearly with the parameter $\taux$ that governs the timescale for the evolution of the effective temperature in Eqn.~\ref{eqn:xdynamics}}.

\begin{figure}[!t]
\subfloat{%
\begin{overpic}[width=0.95\columnwidth]{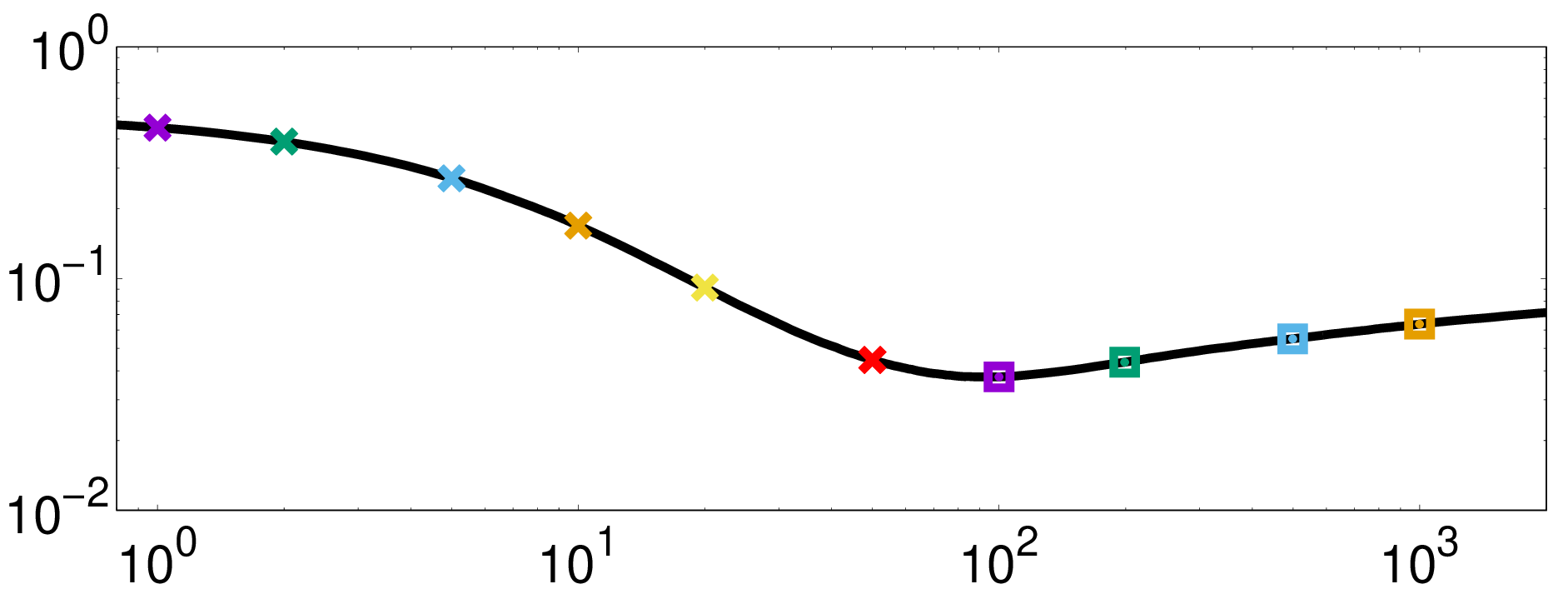}
\put(80,325){\normalsize{a)}}
\put(0,275){\rotatebox{0}{$\mathlarger{\mathlarger{\mathlarger{\mathlarger{\sigma}}}}$}}%
\put(500,-10){$\mathlarger{\mathlarger{\mathlarger{\mathlarger{t}}}}$}%
\end{overpic}
}\hfill\vspace{0cm}
\subfloat{%
\begin{overpic}[width=0.55\columnwidth]{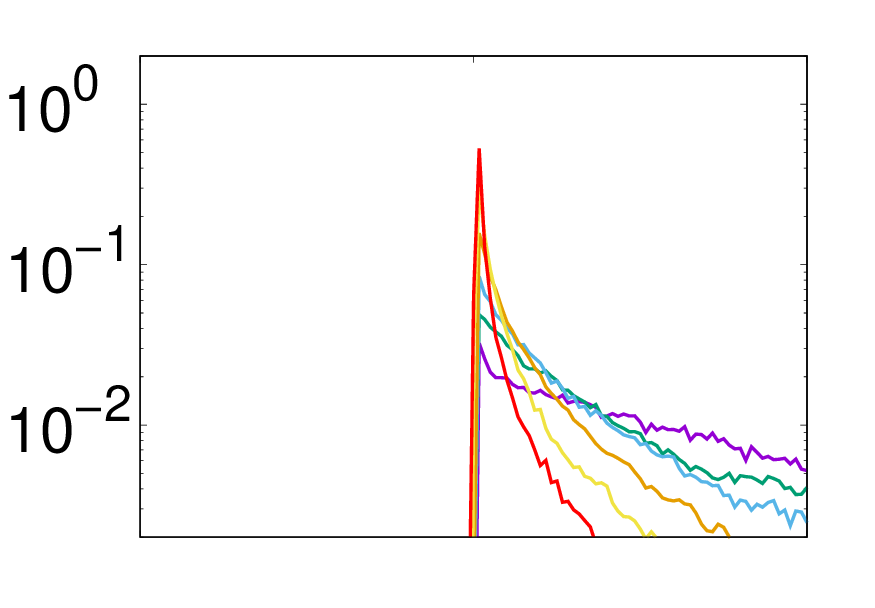}
\put(180,570){\normalsize{b)}}
\put(-20,470) {$\mathlarger{\mathlarger{P(l_h)}}$}%
\linethickness{2pt}
\put(570,200){\vector(0,1){50}}
\put(570,200){\line(0,-1){100}}
\put(800,140){\vector(0,-1){50}}
\put(800,140){\line(0,1){100}}
\put(400,650) {$\mathlarger{t=1-50}$}%
\end{overpic}
}%
\hspace{-1.2cm}
\subfloat{%
\begin{overpic}[width=0.55\columnwidth]{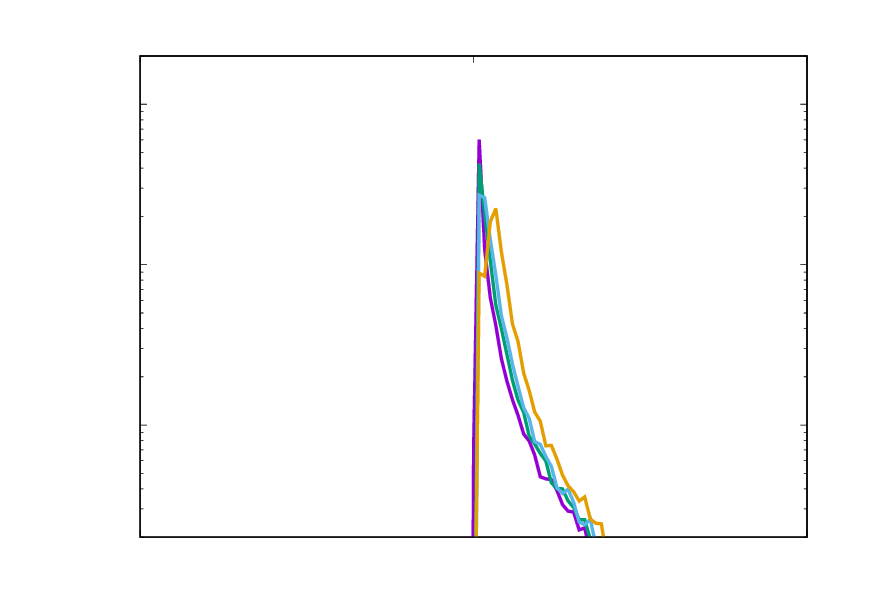}
\put(180,570){\normalsize{c)}}
\linethickness{2pt}
\put(650,470){\vector(1,0){50}}
\put(650,470){\line(-1,0){110}}
\put(400,650) {$\mathlarger{t=100-1000}$}%
\end{overpic}
}%
\vspace{-0.4cm}
\subfloat{%
\begin{overpic}[width=0.55\columnwidth]{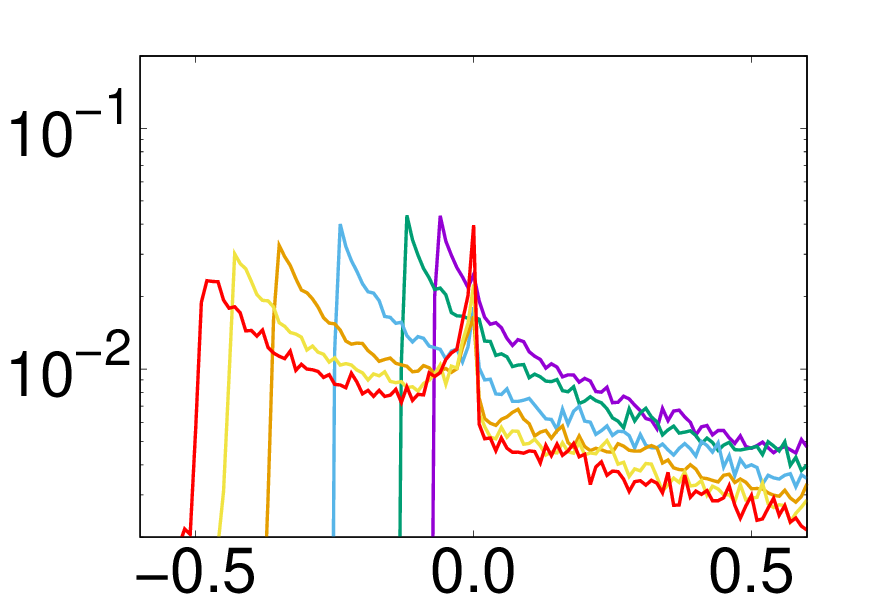}
\put(180,570){\normalsize{d)}}
\put(-20,400) {$\mathlarger{\mathlarger{P(l_u)}}$}%
\linethickness{2pt}
\put(250,470){\vector(-1,0){50}}
\put(250,470){\line(1,0){250}}
\put(700,-10){$\mathlarger{\mathlarger{{{l}}}}$}%
\end{overpic}
}
\hspace{-1.2cm}
\subfloat{%
\begin{overpic}[width=0.55\columnwidth]{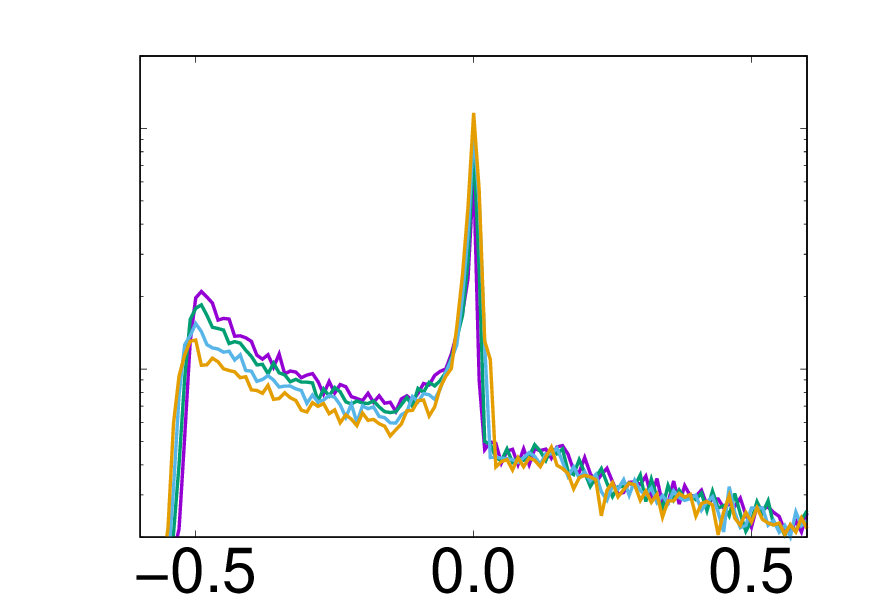}
\put(180,570){\normalsize{e)}}
\linethickness{2pt}
\put(510,490){\vector(0,1){50}}
\put(510,490){\line(0,-1){110}}
\put(240,180){\vector(0,-1){50}}
\put(240,180){\line(0,1){110}}
\put(700,-10){$\mathlarger{\mathlarger{{{l}}}}$}%
\end{overpic}
}\hfill

\caption{\smf{{\bf a)} Non-monotonic stress decay as a function of time $t$ after flow cessation. The distributions of local stresses are shown in panels b+d) at times denoted by crosses during the initial stress decrease in a). Panels c+e) show the distributions during the later stress increase, at times denoted by squares in a). Panels b+c)  pertain to the region previously (during flow) occupied by the high shear band. Panels (d+e) pertain to the region previously occupied by the low shear band.  {\bf b)} In the liquid-like region, elements with positive values of $l$ plastically yield at early times, causing the stress to relax. {\bf c)} At longer times in this region, the distribution shifts slightly to higher local strain values. {\bf d)} In the solid-like region, the distribution shifts elastically leftwards to lower local strain values during the initial stress relaxation. {\bf e)} At longer times in the solid-like region, elements with negative values of $l$ plastically yield. It is this yielding of negative local stresses that results in the increase in macroscopic stress.} Parameter values are $\aparam=6, \taux=100, \gdotbar=0.08$ and $\lambda=0.005$.}
\label{fig:mechanism}
\end{figure}

{\it Mechanistic understanding ---} We now explain the physical mechanism of this phenomenon. \smf{We do so with reference to the distribution of local strains $l$, or equivalently of local stresses $kl$ (recall $k=1$). This is plotted at several times after flow cessation in the region previously  (during flow) occupied by the high shear band  in Figs.~\ref{fig:mechanism}b,c); and in the region previously occupied by the low shear band in Figs.~\ref{fig:mechanism}d,e).  We have checked that the width of these two regions stays relatively constant during the stress relaxation post-cessation.}

Consider a shear banded state before switch-off, with averages over elements in the high shear band and unsheared band denoted $\langle \cdot\rangle_{\rm h}$ 
 and $\langle \cdot \rangle_{\rm u}$. In the high shear band, the strong shear will have fluidised the material, with elements occupying shallow traps of high average yielding rate $\langle r \rangle_{\rm h}$. In the unsheared band the material will be solid-like, elements having aged into deep traps with low $\langle r \rangle_{\rm u}$. Force balance however demands a uniform shear stress, $\sigma=\langle l\rangle_{\rm h}=\langle l\rangle_{\rm u}$. 

Consider now the system's evolution after the shear is switched off. The rate of change of stress $\sdot=\dot{\langle l\rangle_{i}}=\gdot_{i} -\langle l r\rangle_{i}$, with a rate of elastic stress evolution $\gdot$, and of plastic stress relaxation $\langle l r\rangle$. This applies separately in the fluidised and solid-like regions, previously  occupied by the high and low shear bands for which we keep the labels $i={\rm h}, {\rm u}$ even after switch-off. Indeed, each region has a non-zero shear rate $\gdot_{\rm h,u}$ for some time after the shear is switched off, but now with zero average, $\gdotbar=0$. 

In the fluidised region, the rate of plastic stress relaxation $\langle l r\rangle_{\rm h}$ will be high. \smf{Accordingly, the initially broad distribution of values of $l$ (purple line) quickly relaxes to a much narrower distribution focused nearer the origin (red line), as shown in Fig.~\ref{fig:mechanism}b)}. This results in a \smf{rapid} stress decay, giving negative $\sdot$. Force balance demands the same negative $\sdot$ in the solid-like region, in which $\langle l r\rangle_{\rm u}$ is small. A stress decay (negative $\sdot$) thus demands a negative $\gdot_{\rm u}$, i.e., a backwards elastic recoil of the solid-like region, as the material unloads.  This elastic recoil advects the local strains of all elements in that solid-like region to lower values (Fig.~\ref{fig:mechanism}d)  -- including into the negative range of $l$. Over a subsequent, slower timescale, these elements then plastically yield (Fig.~\ref{fig:mechanism}e). In particular, those elements that yield from negative $l$ values and reset their $l$ values to zero contribute a negative rate of stress relaxation $\langle l r\rangle_{\rm u}$, giving a positive contribution to the stress rate of change, $\sdot=\gdot_{\rm u} -\langle l r\rangle_{\rm u}$. {\em It is this plastic yielding of elements with negative $l$ values that can give rise to the observed increase in the macroscopic stress.}  It does so for systems in which the shear bands prior to flow cessation had a sufficiently large difference in effective temperature, and in which the decay time for relaxation  after flow cessation is sufficiently slow. This results in a separation of timescales between (first) the initially fast plastic yielding of positive local stresses from the fluidised band, which causes the initial rapid decay of the macroscopic stress; and (second) the later plastic yielding of negative local stresses from the solid band, which causes the slow rise in macroscopic stress.

{\it Conclusions ---}  We have shown that non-monotonic stress relaxation after flow cessation can originate from an interplay between shear bands present prior to flow cessation. After flow cessation, a rapid plastic stress decay in the fluidised band couples via force balance to give an elastic recoil in the solid-like band. This causes the distribution of local stresses in the solid band to shift to include negative values. The subsequent plastic yielding of these negatively stressed elements, on a longer timescale due to the lower effective temperature of this band, then contributes a slow increase to the macroscopic stress. These two distinct regimes in the stress evolution are strongly reminiscent of observations in boehmite gels~\cite{Sudreau2022} and self-associated organo/supramolecular gels~\cite{Hendricks2019}. We have considered throughout a non-zero noise temperature, for which the terminal time for stress relaxation is large but finite. It would be interesting in future work to consider model predictions at zero temperature, potentially allowing permanent residual stresses, with their important implications for material performance.

{\it Acknowledgements ---} This project has received funding from the European Research Council (ERC) under the European Union's Horizon 2020 research and innovation programme (grant agreement No. 885146). The authors thank Edwina Yeo and Beth Bromley for discussions.

\bibliography{nonmonotonic.bib}

\providecommand{\noopsort}[1]{}\providecommand{\singleletter}[1]{#1}%
\begin{thebibliography}{34}%
\makeatletter
\providecommand \@ifxundefined [1]{%
 \@ifx{#1\undefined}
}%
\providecommand \@ifnum [1]{%
 \ifnum #1\expandafter \@firstoftwo
 \else \expandafter \@secondoftwo
 \fi
}%
\providecommand \@ifx [1]{%
 \ifx #1\expandafter \@firstoftwo
 \else \expandafter \@secondoftwo
 \fi
}%
\providecommand \natexlab [1]{#1}%
\providecommand \enquote  [1]{``#1''}%
\providecommand \bibnamefont  [1]{#1}%
\providecommand \bibfnamefont [1]{#1}%
\providecommand \citenamefont [1]{#1}%
\providecommand \href@noop [0]{\@secondoftwo}%
\providecommand \href [0]{\begingroup \@sanitize@url \@href}%
\providecommand \@href[1]{\@@startlink{#1}\@@href}%
\providecommand \@@href[1]{\endgroup#1\@@endlink}%
\providecommand \@sanitize@url [0]{\catcode `\\12\catcode `\$12\catcode `\&12\catcode `\#12\catcode `\^12\catcode `\_12\catcode `\%12\relax}%
\providecommand \@@startlink[1]{}%
\providecommand \@@endlink[0]{}%
\providecommand \url  [0]{\begingroup\@sanitize@url \@url }%
\providecommand \@url [1]{\endgroup\@href {#1}{\urlprefix }}%
\providecommand \urlprefix  [0]{URL }%
\providecommand \Eprint [0]{\href }%
\providecommand \doibase [0]{https://doi.org/}%
\providecommand \selectlanguage [0]{\@gobble}%
\providecommand \bibinfo  [0]{\@secondoftwo}%
\providecommand \bibfield  [0]{\@secondoftwo}%
\providecommand \translation [1]{[#1]}%
\providecommand \BibitemOpen [0]{}%
\providecommand \bibitemStop [0]{}%
\providecommand \bibitemNoStop [0]{.\EOS\space}%
\providecommand \EOS [0]{\spacefactor3000\relax}%
\providecommand \BibitemShut  [1]{\csname bibitem#1\endcsname}%
\let\auto@bib@innerbib\@empty
\bibitem [{\citenamefont {Larson}(1999)}]{larson1999structure}%
  \BibitemOpen
  \bibfield  {author} {\bibinfo {author} {\bibfnamefont {R.~G.}\ \bibnamefont {Larson}},\ }\href@noop {} {\emph {\bibinfo {title} {The structure and rheology of complex fluids}}}\ (\bibinfo  {publisher} {OUP},\ \bibinfo {year} {1999})\BibitemShut {NoStop}%
\bibitem [{\citenamefont {Berthier}\ and\ \citenamefont {Biroli}(2011)}]{Berthier2011}%
  \BibitemOpen
  \bibfield  {author} {\bibinfo {author} {\bibfnamefont {L.}~\bibnamefont {Berthier}}\ and\ \bibinfo {author} {\bibfnamefont {G.}~\bibnamefont {Biroli}},\ }\bibfield  {title} {\bibinfo {title} {Theoretical perspective on the glass transition and amorphous materials},\ }\href@noop {} {\bibfield  {journal} {\bibinfo  {journal} {Reviews of modern physics}\ }\textbf {\bibinfo {volume} {83}},\ \bibinfo {pages} {587} (\bibinfo {year} {2011})}\BibitemShut {NoStop}%
\bibitem [{\citenamefont {Nicolas}\ \emph {et~al.}(2018)\citenamefont {Nicolas}, \citenamefont {Ferrero}, \citenamefont {Martens},\ and\ \citenamefont {Barrat}}]{Nicolas2018}%
  \BibitemOpen
  \bibfield  {author} {\bibinfo {author} {\bibfnamefont {A.}~\bibnamefont {Nicolas}}, \bibinfo {author} {\bibfnamefont {E.~E.}\ \bibnamefont {Ferrero}}, \bibinfo {author} {\bibfnamefont {K.}~\bibnamefont {Martens}},\ and\ \bibinfo {author} {\bibfnamefont {J.-L.}\ \bibnamefont {Barrat}},\ }\bibfield  {title} {\bibinfo {title} {Deformation and flow of amorphous solids: Insights from elastoplastic models},\ }\href {https://doi.org/10.1103/RevModPhys.90.045006} {\bibfield  {journal} {\bibinfo  {journal} {Rev. Mod. Phys.}\ }\textbf {\bibinfo {volume} {90}},\ \bibinfo {pages} {045006} (\bibinfo {year} {2018})}\BibitemShut {NoStop}%
\bibitem [{\citenamefont {Bonn}\ \emph {et~al.}(2017)\citenamefont {Bonn}, \citenamefont {Denn}, \citenamefont {Berthier}, \citenamefont {Divoux},\ and\ \citenamefont {Manneville}}]{bonn2017yield}%
  \BibitemOpen
  \bibfield  {author} {\bibinfo {author} {\bibfnamefont {D.}~\bibnamefont {Bonn}}, \bibinfo {author} {\bibfnamefont {M.~M.}\ \bibnamefont {Denn}}, \bibinfo {author} {\bibfnamefont {L.}~\bibnamefont {Berthier}}, \bibinfo {author} {\bibfnamefont {T.}~\bibnamefont {Divoux}},\ and\ \bibinfo {author} {\bibfnamefont {S.}~\bibnamefont {Manneville}},\ }\bibfield  {title} {\bibinfo {title} {Yield stress materials in soft condensed matter},\ }\href@noop {} {\bibfield  {journal} {\bibinfo  {journal} {Reviews of Modern Physics}\ }\textbf {\bibinfo {volume} {89}},\ \bibinfo {pages} {035005} (\bibinfo {year} {2017})}\BibitemShut {NoStop}%
\bibitem [{\citenamefont {Divoux}\ \emph {et~al.}(2023)\citenamefont {Divoux}, \citenamefont {Agoritsas}, \citenamefont {Aime}, \citenamefont {Barentin}, \citenamefont {Barrat}, \citenamefont {Benzi}, \citenamefont {Berthier}, \citenamefont {Bi}, \citenamefont {Biroli}, \citenamefont {Bonn} \emph {et~al.}}]{divoux2023ductile}%
  \BibitemOpen
  \bibfield  {author} {\bibinfo {author} {\bibfnamefont {T.}~\bibnamefont {Divoux}}, \bibinfo {author} {\bibfnamefont {E.}~\bibnamefont {Agoritsas}}, \bibinfo {author} {\bibfnamefont {S.}~\bibnamefont {Aime}}, \bibinfo {author} {\bibfnamefont {C.}~\bibnamefont {Barentin}}, \bibinfo {author} {\bibfnamefont {J.-L.}\ \bibnamefont {Barrat}}, \bibinfo {author} {\bibfnamefont {R.}~\bibnamefont {Benzi}}, \bibinfo {author} {\bibfnamefont {L.}~\bibnamefont {Berthier}}, \bibinfo {author} {\bibfnamefont {D.}~\bibnamefont {Bi}}, \bibinfo {author} {\bibfnamefont {G.}~\bibnamefont {Biroli}}, \bibinfo {author} {\bibfnamefont {D.}~\bibnamefont {Bonn}}, \emph {et~al.},\ }\bibfield  {title} {\bibinfo {title} {Ductile-to-brittle transition and yielding in soft amorphous materials: perspectives and open questions},\ }\href@noop {} {\bibfield  {journal} {\bibinfo  {journal} {arXiv preprint arXiv:2312.14278}\ } (\bibinfo {year} {2023})}\BibitemShut {NoStop}%
\bibitem [{\citenamefont {Ballauff}\ \emph {et~al.}(2013)\citenamefont {Ballauff}, \citenamefont {Brader}, \citenamefont {Egelhaaf}, \citenamefont {Fuchs}, \citenamefont {Horbach}, \citenamefont {Koumakis}, \citenamefont {Kr\"uger}, \citenamefont {Laurati}, \citenamefont {Mutch}, \citenamefont {Petekidis}, \citenamefont {Siebenb\"urger}, \citenamefont {Voigtmann},\ and\ \citenamefont {Zausch}}]{Ballauff2013}%
  \BibitemOpen
  \bibfield  {author} {\bibinfo {author} {\bibfnamefont {M.}~\bibnamefont {Ballauff}}, \bibinfo {author} {\bibfnamefont {J.~M.}\ \bibnamefont {Brader}}, \bibinfo {author} {\bibfnamefont {S.~U.}\ \bibnamefont {Egelhaaf}}, \bibinfo {author} {\bibfnamefont {M.}~\bibnamefont {Fuchs}}, \bibinfo {author} {\bibfnamefont {J.}~\bibnamefont {Horbach}}, \bibinfo {author} {\bibfnamefont {N.}~\bibnamefont {Koumakis}}, \bibinfo {author} {\bibfnamefont {M.}~\bibnamefont {Kr\"uger}}, \bibinfo {author} {\bibfnamefont {M.}~\bibnamefont {Laurati}}, \bibinfo {author} {\bibfnamefont {K.~J.}\ \bibnamefont {Mutch}}, \bibinfo {author} {\bibfnamefont {G.}~\bibnamefont {Petekidis}}, \bibinfo {author} {\bibfnamefont {M.}~\bibnamefont {Siebenb\"urger}}, \bibinfo {author} {\bibfnamefont {T.}~\bibnamefont {Voigtmann}},\ and\ \bibinfo {author} {\bibfnamefont {J.}~\bibnamefont {Zausch}},\ }\bibfield  {title} {\bibinfo {title} {Residual stresses in glasses},\ }\href {https://doi.org/10.1103/PhysRevLett.110.215701} {\bibfield  {journal}
  {\bibinfo  {journal} {Phys. Rev. Lett.}\ }\textbf {\bibinfo {volume} {110}},\ \bibinfo {pages} {215701} (\bibinfo {year} {2013})}\BibitemShut {NoStop}%
\bibitem [{\citenamefont {Moghimi}\ \emph {et~al.}(2017)\citenamefont {Moghimi}, \citenamefont {Jacob},\ and\ \citenamefont {Petekidis}}]{Moghimi2017}%
  \BibitemOpen
  \bibfield  {author} {\bibinfo {author} {\bibfnamefont {E.}~\bibnamefont {Moghimi}}, \bibinfo {author} {\bibfnamefont {A.~R.}\ \bibnamefont {Jacob}},\ and\ \bibinfo {author} {\bibfnamefont {G.}~\bibnamefont {Petekidis}},\ }\bibfield  {title} {\bibinfo {title} {Residual stresses in colloidal gels},\ }\href {https://doi.org/10.1039/C7SM01655G} {\bibfield  {journal} {\bibinfo  {journal} {Soft Matter}\ }\textbf {\bibinfo {volume} {13}},\ \bibinfo {pages} {7824} (\bibinfo {year} {2017})}\BibitemShut {NoStop}%
\bibitem [{\citenamefont {Mohan}\ \emph {et~al.}(2013)\citenamefont {Mohan}, \citenamefont {Bonnecaze},\ and\ \citenamefont {Cloitre}}]{Mohan2013}%
  \BibitemOpen
  \bibfield  {author} {\bibinfo {author} {\bibfnamefont {L.}~\bibnamefont {Mohan}}, \bibinfo {author} {\bibfnamefont {R.~T.}\ \bibnamefont {Bonnecaze}},\ and\ \bibinfo {author} {\bibfnamefont {M.}~\bibnamefont {Cloitre}},\ }\bibfield  {title} {\bibinfo {title} {Microscopic origin of internal stresses in jammed soft particle suspensions},\ }\href {https://doi.org/10.1103/PhysRevLett.111.268301} {\bibfield  {journal} {\bibinfo  {journal} {Phys. Rev. Lett.}\ }\textbf {\bibinfo {volume} {111}},\ \bibinfo {pages} {268301} (\bibinfo {year} {2013})}\BibitemShut {NoStop}%
\bibitem [{\citenamefont {Vasisht}\ \emph {et~al.}(2022)\citenamefont {Vasisht}, \citenamefont {Chaudhuri},\ and\ \citenamefont {Martens}}]{Vasisht2022}%
  \BibitemOpen
  \bibfield  {author} {\bibinfo {author} {\bibfnamefont {V.~V.}\ \bibnamefont {Vasisht}}, \bibinfo {author} {\bibfnamefont {P.}~\bibnamefont {Chaudhuri}},\ and\ \bibinfo {author} {\bibfnamefont {K.}~\bibnamefont {Martens}},\ }\bibfield  {title} {\bibinfo {title} {Residual stress in athermal soft disordered solids: insights from microscopic and mesoscale models},\ }\href@noop {} {\bibfield  {journal} {\bibinfo  {journal} {Soft Matter}\ }\textbf {\bibinfo {volume} {18}},\ \bibinfo {pages} {6426} (\bibinfo {year} {2022})}\BibitemShut {NoStop}%
\bibitem [{\citenamefont {Osuji}\ \emph {et~al.}(2008)\citenamefont {Osuji}, \citenamefont {Kim},\ and\ \citenamefont {Weitz}}]{osuji2008shear}%
  \BibitemOpen
  \bibfield  {author} {\bibinfo {author} {\bibfnamefont {C.~O.}\ \bibnamefont {Osuji}}, \bibinfo {author} {\bibfnamefont {C.}~\bibnamefont {Kim}},\ and\ \bibinfo {author} {\bibfnamefont {D.~A.}\ \bibnamefont {Weitz}},\ }\bibfield  {title} {\bibinfo {title} {Shear thickening and scaling of the elastic modulus in a fractal colloidal system with attractive interactions},\ }\href@noop {} {\bibfield  {journal} {\bibinfo  {journal} {Physical Review E}\ }\textbf {\bibinfo {volume} {77}},\ \bibinfo {pages} {060402(R)} (\bibinfo {year} {2008})}\BibitemShut {NoStop}%
\bibitem [{\citenamefont {Negi}\ and\ \citenamefont {Osuji}(2010)}]{negi2010physical}%
  \BibitemOpen
  \bibfield  {author} {\bibinfo {author} {\bibfnamefont {A.~S.}\ \bibnamefont {Negi}}\ and\ \bibinfo {author} {\bibfnamefont {C.~O.}\ \bibnamefont {Osuji}},\ }\bibfield  {title} {\bibinfo {title} {Physical aging and relaxation of residual stresses in a colloidal glass following flow cessation},\ }\href@noop {} {\bibfield  {journal} {\bibinfo  {journal} {Journal of Rheology}\ }\textbf {\bibinfo {volume} {54}},\ \bibinfo {pages} {943} (\bibinfo {year} {2010})}\BibitemShut {NoStop}%
\bibitem [{\citenamefont {Fritschi}\ \emph {et~al.}(2014)\citenamefont {Fritschi}, \citenamefont {Fuchs},\ and\ \citenamefont {Voigtmann}}]{fritschi2014mode}%
  \BibitemOpen
  \bibfield  {author} {\bibinfo {author} {\bibfnamefont {S.}~\bibnamefont {Fritschi}}, \bibinfo {author} {\bibfnamefont {M.}~\bibnamefont {Fuchs}},\ and\ \bibinfo {author} {\bibfnamefont {T.}~\bibnamefont {Voigtmann}},\ }\bibfield  {title} {\bibinfo {title} {Mode-coupling analysis of residual stresses in colloidal glasses},\ }\href@noop {} {\bibfield  {journal} {\bibinfo  {journal} {Soft matter}\ }\textbf {\bibinfo {volume} {10}},\ \bibinfo {pages} {4822} (\bibinfo {year} {2014})}\BibitemShut {NoStop}%
\bibitem [{\citenamefont {Pashine}\ \emph {et~al.}(2019)\citenamefont {Pashine}, \citenamefont {Hexner}, \citenamefont {Liu},\ and\ \citenamefont {Nagel}}]{Pashine2019}%
  \BibitemOpen
  \bibfield  {author} {\bibinfo {author} {\bibfnamefont {N.}~\bibnamefont {Pashine}}, \bibinfo {author} {\bibfnamefont {D.}~\bibnamefont {Hexner}}, \bibinfo {author} {\bibfnamefont {A.~J.}\ \bibnamefont {Liu}},\ and\ \bibinfo {author} {\bibfnamefont {S.~R.}\ \bibnamefont {Nagel}},\ }\bibfield  {title} {\bibinfo {title} {Directed aging, memory, and nature’s greed},\ }\href@noop {} {\bibfield  {journal} {\bibinfo  {journal} {Science advances}\ }\textbf {\bibinfo {volume} {5}},\ \bibinfo {pages} {eaax4215} (\bibinfo {year} {2019})}\BibitemShut {NoStop}%
\bibitem [{\citenamefont {Keim}\ and\ \citenamefont {Arratia}(2014)}]{keim2014mechanical}%
  \BibitemOpen
  \bibfield  {author} {\bibinfo {author} {\bibfnamefont {N.~C.}\ \bibnamefont {Keim}}\ and\ \bibinfo {author} {\bibfnamefont {P.~E.}\ \bibnamefont {Arratia}},\ }\bibfield  {title} {\bibinfo {title} {Mechanical and microscopic properties of the reversible plastic regime in a 2d jammed material},\ }\href@noop {} {\bibfield  {journal} {\bibinfo  {journal} {Physical review letters}\ }\textbf {\bibinfo {volume} {112}},\ \bibinfo {pages} {028302} (\bibinfo {year} {2014})}\BibitemShut {NoStop}%
\bibitem [{\citenamefont {Mukherji}\ \emph {et~al.}(2019)\citenamefont {Mukherji}, \citenamefont {Kandula}, \citenamefont {Sood},\ and\ \citenamefont {Ganapathy}}]{mukherji2019strength}%
  \BibitemOpen
  \bibfield  {author} {\bibinfo {author} {\bibfnamefont {S.}~\bibnamefont {Mukherji}}, \bibinfo {author} {\bibfnamefont {N.}~\bibnamefont {Kandula}}, \bibinfo {author} {\bibfnamefont {A.~K.}\ \bibnamefont {Sood}},\ and\ \bibinfo {author} {\bibfnamefont {R.}~\bibnamefont {Ganapathy}},\ }\bibfield  {title} {\bibinfo {title} {Strength of mechanical memories is maximal at the yield point of a soft glass},\ }\href@noop {} {\bibfield  {journal} {\bibinfo  {journal} {Physical review letters}\ }\textbf {\bibinfo {volume} {122}},\ \bibinfo {pages} {158001} (\bibinfo {year} {2019})}\BibitemShut {NoStop}%
\bibitem [{\citenamefont {Keim}\ \emph {et~al.}(2019)\citenamefont {Keim}, \citenamefont {Paulsen}, \citenamefont {Zeravcic}, \citenamefont {Sastry},\ and\ \citenamefont {Nagel}}]{keim2019memory}%
  \BibitemOpen
  \bibfield  {author} {\bibinfo {author} {\bibfnamefont {N.~C.}\ \bibnamefont {Keim}}, \bibinfo {author} {\bibfnamefont {J.~D.}\ \bibnamefont {Paulsen}}, \bibinfo {author} {\bibfnamefont {Z.}~\bibnamefont {Zeravcic}}, \bibinfo {author} {\bibfnamefont {S.}~\bibnamefont {Sastry}},\ and\ \bibinfo {author} {\bibfnamefont {S.~R.}\ \bibnamefont {Nagel}},\ }\bibfield  {title} {\bibinfo {title} {Memory formation in matter},\ }\href@noop {} {\bibfield  {journal} {\bibinfo  {journal} {Reviews of Modern Physics}\ }\textbf {\bibinfo {volume} {91}},\ \bibinfo {pages} {035002} (\bibinfo {year} {2019})}\BibitemShut {NoStop}%
\bibitem [{\citenamefont {Hirobe}\ \emph {et~al.}(2021)\citenamefont {Hirobe}, \citenamefont {Imakita}, \citenamefont {Aizawa}, \citenamefont {Kato}, \citenamefont {Urata},\ and\ \citenamefont {Oguni}}]{hirobe2021simulation}%
  \BibitemOpen
  \bibfield  {author} {\bibinfo {author} {\bibfnamefont {S.}~\bibnamefont {Hirobe}}, \bibinfo {author} {\bibfnamefont {K.}~\bibnamefont {Imakita}}, \bibinfo {author} {\bibfnamefont {H.}~\bibnamefont {Aizawa}}, \bibinfo {author} {\bibfnamefont {Y.}~\bibnamefont {Kato}}, \bibinfo {author} {\bibfnamefont {S.}~\bibnamefont {Urata}},\ and\ \bibinfo {author} {\bibfnamefont {K.}~\bibnamefont {Oguni}},\ }\bibfield  {title} {\bibinfo {title} {Simulation of catastrophic failure in a residual stress field},\ }\href@noop {} {\bibfield  {journal} {\bibinfo  {journal} {Physical Review Letters}\ }\textbf {\bibinfo {volume} {127}},\ \bibinfo {pages} {064301} (\bibinfo {year} {2021})}\BibitemShut {NoStop}%
\bibitem [{\citenamefont {Kooij}\ \emph {et~al.}(2021)\citenamefont {Kooij}, \citenamefont {van Dalen}, \citenamefont {Molinari},\ and\ \citenamefont {Bonn}}]{kooij2021explosive}%
  \BibitemOpen
  \bibfield  {author} {\bibinfo {author} {\bibfnamefont {S.}~\bibnamefont {Kooij}}, \bibinfo {author} {\bibfnamefont {G.}~\bibnamefont {van Dalen}}, \bibinfo {author} {\bibfnamefont {J.-F.}\ \bibnamefont {Molinari}},\ and\ \bibinfo {author} {\bibfnamefont {D.}~\bibnamefont {Bonn}},\ }\bibfield  {title} {\bibinfo {title} {Explosive fragmentation of prince rupert’s drops leads to well-defined fragment sizes},\ }\href@noop {} {\bibfield  {journal} {\bibinfo  {journal} {Nature communications}\ }\textbf {\bibinfo {volume} {12}},\ \bibinfo {pages} {2521} (\bibinfo {year} {2021})}\BibitemShut {NoStop}%
\bibitem [{\citenamefont {Withers}(2007)}]{withers2007residual}%
  \BibitemOpen
  \bibfield  {author} {\bibinfo {author} {\bibfnamefont {P.}~\bibnamefont {Withers}},\ }\bibfield  {title} {\bibinfo {title} {Residual stress and its role in failure},\ }\href@noop {} {\bibfield  {journal} {\bibinfo  {journal} {Reports on progress in physics}\ }\textbf {\bibinfo {volume} {70}},\ \bibinfo {pages} {2211} (\bibinfo {year} {2007})}\BibitemShut {NoStop}%
\bibitem [{\citenamefont {Sudreau}\ \emph {et~al.}(2022)\citenamefont {Sudreau}, \citenamefont {Auxois}, \citenamefont {Servel}, \citenamefont {L\'ecolier}, \citenamefont {Manneville},\ and\ \citenamefont {Divoux}}]{Sudreau2022}%
  \BibitemOpen
  \bibfield  {author} {\bibinfo {author} {\bibfnamefont {I.}~\bibnamefont {Sudreau}}, \bibinfo {author} {\bibfnamefont {M.}~\bibnamefont {Auxois}}, \bibinfo {author} {\bibfnamefont {M.}~\bibnamefont {Servel}}, \bibinfo {author} {\bibfnamefont {E.}~\bibnamefont {L\'ecolier}}, \bibinfo {author} {\bibfnamefont {S.}~\bibnamefont {Manneville}},\ and\ \bibinfo {author} {\bibfnamefont {T.}~\bibnamefont {Divoux}},\ }\bibfield  {title} {\bibinfo {title} {Residual stresses and shear-induced overaging in boehmite gels},\ }\href {https://doi.org/10.1103/PhysRevMaterials.6.L042601} {\bibfield  {journal} {\bibinfo  {journal} {Phys. Rev. Mater.}\ }\textbf {\bibinfo {volume} {6}},\ \bibinfo {pages} {L042601} (\bibinfo {year} {2022})}\BibitemShut {NoStop}%
\bibitem [{\citenamefont {Hendricks}\ \emph {et~al.}(2019)\citenamefont {Hendricks}, \citenamefont {Louhichi}, \citenamefont {Metri}, \citenamefont {Fournier}, \citenamefont {Reddy}, \citenamefont {Bouteiller}, \citenamefont {Cloitre}, \citenamefont {Clasen}, \citenamefont {Vlassopoulos},\ and\ \citenamefont {Briels}}]{Hendricks2019}%
  \BibitemOpen
  \bibfield  {author} {\bibinfo {author} {\bibfnamefont {J.}~\bibnamefont {Hendricks}}, \bibinfo {author} {\bibfnamefont {A.}~\bibnamefont {Louhichi}}, \bibinfo {author} {\bibfnamefont {V.}~\bibnamefont {Metri}}, \bibinfo {author} {\bibfnamefont {R.}~\bibnamefont {Fournier}}, \bibinfo {author} {\bibfnamefont {N.}~\bibnamefont {Reddy}}, \bibinfo {author} {\bibfnamefont {L.}~\bibnamefont {Bouteiller}}, \bibinfo {author} {\bibfnamefont {M.}~\bibnamefont {Cloitre}}, \bibinfo {author} {\bibfnamefont {C.}~\bibnamefont {Clasen}}, \bibinfo {author} {\bibfnamefont {D.}~\bibnamefont {Vlassopoulos}},\ and\ \bibinfo {author} {\bibfnamefont {W.~J.}\ \bibnamefont {Briels}},\ }\bibfield  {title} {\bibinfo {title} {Nonmonotonic stress relaxation after cessation of steady shear flow in supramolecular assemblies},\ }\href {https://doi.org/10.1103/PhysRevLett.123.218003} {\bibfield  {journal} {\bibinfo  {journal} {Phys. Rev. Lett.}\ }\textbf {\bibinfo {volume} {123}},\ \bibinfo {pages} {218003} (\bibinfo {year}
  {2019})}\BibitemShut {NoStop}%
\bibitem [{\citenamefont {Murphy}\ \emph {et~al.}(2020)\citenamefont {Murphy}, \citenamefont {Kruppe},\ and\ \citenamefont {Jaeger}}]{Murphy2020}%
  \BibitemOpen
  \bibfield  {author} {\bibinfo {author} {\bibfnamefont {K.~A.}\ \bibnamefont {Murphy}}, \bibinfo {author} {\bibfnamefont {J.~W.}\ \bibnamefont {Kruppe}},\ and\ \bibinfo {author} {\bibfnamefont {H.~M.}\ \bibnamefont {Jaeger}},\ }\bibfield  {title} {\bibinfo {title} {Memory in nonmonotonic stress relaxation of a granular system},\ }\href@noop {} {\bibfield  {journal} {\bibinfo  {journal} {Physical Review Letters}\ }\textbf {\bibinfo {volume} {124}},\ \bibinfo {pages} {168002} (\bibinfo {year} {2020})}\BibitemShut {NoStop}%
\bibitem [{\citenamefont {Lahini}\ \emph {et~al.}(2017)\citenamefont {Lahini}, \citenamefont {Gottesman}, \citenamefont {Amir},\ and\ \citenamefont {Rubinstein}}]{Lahini2017}%
  \BibitemOpen
  \bibfield  {author} {\bibinfo {author} {\bibfnamefont {Y.}~\bibnamefont {Lahini}}, \bibinfo {author} {\bibfnamefont {O.}~\bibnamefont {Gottesman}}, \bibinfo {author} {\bibfnamefont {A.}~\bibnamefont {Amir}},\ and\ \bibinfo {author} {\bibfnamefont {S.~M.}\ \bibnamefont {Rubinstein}},\ }\bibfield  {title} {\bibinfo {title} {Nonmonotonic aging and memory retention in disordered mechanical systems},\ }\href@noop {} {\bibfield  {journal} {\bibinfo  {journal} {Physical review letters}\ }\textbf {\bibinfo {volume} {118}},\ \bibinfo {pages} {085501} (\bibinfo {year} {2017})}\BibitemShut {NoStop}%
\bibitem [{\citenamefont {Mandal}\ \emph {et~al.}(2021)\citenamefont {Mandal}, \citenamefont {Tapias},\ and\ \citenamefont {Sollich}}]{mandal2021memory}%
  \BibitemOpen
  \bibfield  {author} {\bibinfo {author} {\bibfnamefont {R.}~\bibnamefont {Mandal}}, \bibinfo {author} {\bibfnamefont {D.}~\bibnamefont {Tapias}},\ and\ \bibinfo {author} {\bibfnamefont {P.}~\bibnamefont {Sollich}},\ }\bibfield  {title} {\bibinfo {title} {Memory in non-monotonic stress response of an athermal disordered solid},\ }\href@noop {} {\bibfield  {journal} {\bibinfo  {journal} {Physical Review Research}\ }\textbf {\bibinfo {volume} {3}},\ \bibinfo {pages} {043153} (\bibinfo {year} {2021})}\BibitemShut {NoStop}%
\bibitem [{\citenamefont {Joshi}(2022)}]{joshi2022thixotropy}%
  \BibitemOpen
  \bibfield  {author} {\bibinfo {author} {\bibfnamefont {Y.~M.}\ \bibnamefont {Joshi}},\ }\bibfield  {title} {\bibinfo {title} {Thixotropy, nonmonotonic stress relaxation, and the second law of thermodynamics},\ }\href@noop {} {\bibfield  {journal} {\bibinfo  {journal} {Journal of Rheology}\ }\textbf {\bibinfo {volume} {66}},\ \bibinfo {pages} {111} (\bibinfo {year} {2022})}\BibitemShut {NoStop}%
\bibitem [{\citenamefont {Larson}(2015)}]{larson2015constitutive}%
  \BibitemOpen
  \bibfield  {author} {\bibinfo {author} {\bibfnamefont {R.}~\bibnamefont {Larson}},\ }\bibfield  {title} {\bibinfo {title} {Constitutive equations for thixotropic fluids},\ }\href@noop {} {\bibfield  {journal} {\bibinfo  {journal} {Journal of Rheology}\ }\textbf {\bibinfo {volume} {59}},\ \bibinfo {pages} {595} (\bibinfo {year} {2015})}\BibitemShut {NoStop}%
\bibitem [{\citenamefont {Choi}\ \emph {et~al.}(2021)\citenamefont {Choi}, \citenamefont {Armstrong},\ and\ \citenamefont {Rogers}}]{choi2021role}%
  \BibitemOpen
  \bibfield  {author} {\bibinfo {author} {\bibfnamefont {J.}~\bibnamefont {Choi}}, \bibinfo {author} {\bibfnamefont {M.}~\bibnamefont {Armstrong}},\ and\ \bibinfo {author} {\bibfnamefont {S.~A.}\ \bibnamefont {Rogers}},\ }\bibfield  {title} {\bibinfo {title} {The role of elasticity in thixotropy: Transient elastic stress during stepwise reduction in shear rate},\ }\href@noop {} {\bibfield  {journal} {\bibinfo  {journal} {Physics of Fluids}\ }\textbf {\bibinfo {volume} {33}} (\bibinfo {year} {2021})}\BibitemShut {NoStop}%
\bibitem [{\citenamefont {Vlassopoulos}(2024)}]{Vlassopolous.personal}%
  \BibitemOpen
  \bibfield  {author} {\bibinfo {author} {\bibfnamefont {D.}~\bibnamefont {Vlassopoulos}},\ }\href@noop {} {}\bibinfo {howpublished} {Personal communication} (\bibinfo {year} {2024})\BibitemShut {NoStop}%
\bibitem [{\citenamefont {Sollich}\ \emph {et~al.}(1997)\citenamefont {Sollich}, \citenamefont {Lequeux}, \citenamefont {H{\'e}braud},\ and\ \citenamefont {Cates}}]{Sollich1997}%
  \BibitemOpen
  \bibfield  {author} {\bibinfo {author} {\bibfnamefont {P.}~\bibnamefont {Sollich}}, \bibinfo {author} {\bibfnamefont {F.}~\bibnamefont {Lequeux}}, \bibinfo {author} {\bibfnamefont {P.}~\bibnamefont {H{\'e}braud}},\ and\ \bibinfo {author} {\bibfnamefont {M.~E.}\ \bibnamefont {Cates}},\ }\bibfield  {title} {\bibinfo {title} {Rheology of soft glassy materials},\ }\href@noop {} {\bibfield  {journal} {\bibinfo  {journal} {Physical review letters}\ }\textbf {\bibinfo {volume} {78}},\ \bibinfo {pages} {2020} (\bibinfo {year} {1997})}\BibitemShut {NoStop}%
\bibitem [{\citenamefont {Sollich}(1998)}]{Sollich1998}%
  \BibitemOpen
  \bibfield  {author} {\bibinfo {author} {\bibfnamefont {P.}~\bibnamefont {Sollich}},\ }\bibfield  {title} {\bibinfo {title} {Rheological constitutive equation for a model of soft glassy materials},\ }\href {https://doi.org/10.1103/PhysRevE.58.738} {\bibfield  {journal} {\bibinfo  {journal} {Phys. Rev. E}\ }\textbf {\bibinfo {volume} {58}},\ \bibinfo {pages} {738} (\bibinfo {year} {1998})}\BibitemShut {NoStop}%
\bibitem [{\citenamefont {{J. P. Bouchaud}}(1992)}]{Bouchaud1992}%
  \BibitemOpen
  \bibfield  {author} {\bibinfo {author} {\bibnamefont {{J. P. Bouchaud}}},\ }\bibfield  {title} {\bibinfo {title} {Weak ergodicity breaking and aging in disordered systems},\ }\href {https://doi.org/10.1051/jp1:1992238} {\bibfield  {journal} {\bibinfo  {journal} {J. Phys. I France}\ }\textbf {\bibinfo {volume} {2}},\ \bibinfo {pages} {1705} (\bibinfo {year} {1992})}\BibitemShut {NoStop}%
\bibitem [{\citenamefont {Fielding}\ \emph {et~al.}(2000)\citenamefont {Fielding}, \citenamefont {Sollich},\ and\ \citenamefont {Cates}}]{Fielding2000}%
  \BibitemOpen
  \bibfield  {author} {\bibinfo {author} {\bibfnamefont {S.~M.}\ \bibnamefont {Fielding}}, \bibinfo {author} {\bibfnamefont {P.}~\bibnamefont {Sollich}},\ and\ \bibinfo {author} {\bibfnamefont {M.~E.}\ \bibnamefont {Cates}},\ }\bibfield  {title} {\bibinfo {title} {Aging and rheology in soft materials},\ }\href@noop {} {\bibfield  {journal} {\bibinfo  {journal} {Journal of Rheology}\ }\textbf {\bibinfo {volume} {44}},\ \bibinfo {pages} {323} (\bibinfo {year} {2000})}\BibitemShut {NoStop}%
\bibitem [{\citenamefont {Fielding}\ \emph {et~al.}(2009)\citenamefont {Fielding}, \citenamefont {Cates},\ and\ \citenamefont {Sollich}}]{Fielding2009}%
  \BibitemOpen
  \bibfield  {author} {\bibinfo {author} {\bibfnamefont {S.~M.}\ \bibnamefont {Fielding}}, \bibinfo {author} {\bibfnamefont {M.~E.}\ \bibnamefont {Cates}},\ and\ \bibinfo {author} {\bibfnamefont {P.}~\bibnamefont {Sollich}},\ }\bibfield  {title} {\bibinfo {title} {Shear banding, aging and noise dynamics in soft glassy materials},\ }\href {https://doi.org/10.1039/b812394m} {\bibfield  {journal} {\bibinfo  {journal} {Soft Matter}\ }\textbf {\bibinfo {volume} {5}},\ \bibinfo {pages} {2378–2382} (\bibinfo {year} {2009})}\BibitemShut {NoStop}%
\bibitem [{\citenamefont {Picard}\ \emph {et~al.}(2004)\citenamefont {Picard}, \citenamefont {Ajdari}, \citenamefont {Lequeux},\ and\ \citenamefont {Bocquet}}]{picard2004elastic}%
  \BibitemOpen
  \bibfield  {author} {\bibinfo {author} {\bibfnamefont {G.}~\bibnamefont {Picard}}, \bibinfo {author} {\bibfnamefont {A.}~\bibnamefont {Ajdari}}, \bibinfo {author} {\bibfnamefont {F.}~\bibnamefont {Lequeux}},\ and\ \bibinfo {author} {\bibfnamefont {L.}~\bibnamefont {Bocquet}},\ }\bibfield  {title} {\bibinfo {title} {Elastic consequences of a single plastic event: A step towards the microscopic modeling of the flow of yield stress fluids},\ }\href@noop {} {\bibfield  {journal} {\bibinfo  {journal} {The European Physical Journal E}\ }\textbf {\bibinfo {volume} {15}},\ \bibinfo {pages} {371} (\bibinfo {year} {2004})}\BibitemShut {NoStop}%
\end{thebibliography}%

\end{document}